\documentclass[letterpaper,10 pt,conference]{ieeeconf}

\IEEEoverridecommandlockouts
\usepackage{amsmath}
\usepackage{amssymb}
\usepackage{graphicx}
\usepackage{epsfig}
\usepackage{algorithm}
\usepackage{algorithmic}
\usepackage{xcolor}

\newtheorem{Lemma}{Lemma}[section]
\newtheorem{Corollary}{Corollary}[section]

\newtheorem{Assumption}{Assumption}[section]
\newtheorem{Remark}{Remark}[section]

\newcommand{\beq}{\begin{equation}}
\newcommand{\beql}[1]{\begin{equation}\label{#1}}
\newcommand{\eeq}{\end{equation}}

\newcommand{\bDelta}{\bf{\Delta}}
\newcommand{\BD}{\mathcal B_{\bDelta}}

\title{\LARGE \bf
LFT modelling and $\mu$-based robust performance analysis \\ of hybrid multi-rate control systems
}

\author{Jean-Marc Biannic$^*$, Cl\'ement Roos and Christelle Cumer
\thanks{DTIS, ONERA, Universit\'e de Toulouse, 31000 Toulouse, France.} 
\thanks{$^*$ Corresponding author : {\tt\small jean-marc.biannic@onera.fr}}%
}

\begin{document}

\maketitle
\thispagestyle{empty}
\pagestyle{empty}

\begin{abstract}
This paper focuses on robust stability and ${\cal H}_\infty$ performance analyses of hybrid continuous/discrete time linear multi-rate control systems in the presence of parametric uncertainties. These affect the continuous-time plant in a rational way which is then modeled as a Linear Fractional Transformation (LFT). Based on a zero-order-hold (ZOH) LFT discretization process at the cost of bounded quantifiable approximations, and then using LFT-preserving down-sampling operations, a single-rate discrete-time closed-loop LFT model is derived. Interestingly, for any step inputs, and any admissible values of the uncertain parameters, the outputs of this model cover those of the initial hybrid multi-rate closed-loop system at every sampling time of the slowest control loop. Such an LFT model, which also captures the discretization errors, can then be used to evaluate both robust stability and guaranteed $\mathcal{H}_\infty$ performance with a $\mu$-based approach. The proposed methodology is illustrated on a realistic and easily reproducible example inspired by the validation of \mbox{multi-rate attitude control systems.}
\end{abstract}

\section{Introduction}
Although a few methods were already proposed over 30 years ago 
\cite{Berg88,alrhmani_auto_1992,voulgaris_SCL_1993,Chen94_hinfMR} but also quite recently \cite{Cimini_TCST_2009} to design multi-rate control systems taking into account the hybrid nature (simultaneously involving continuous and discrete time dynamics) of a closed-loop system, common engineering practice still consists in designing controllers in the continuous-time domain. This is easily explained by the high level of maturity and the numerous tools available in the field of robust continuous-time control, which has gained in popularity in recent years. 

As a result, to shorten time-consuming simulation-based validation campaigns, there is a growing need for advanced and possibly cheap validation methodologies. Their main objective, which is also central in this paper, is to guarantee (at a reasonably low cost) closed-loop stability and performance after digital implementation of control laws often involving several loops operating at different rates. Unsurprisingly, after the early work of Kranc \cite{Kranc_IRE_TAC_1957}, modeling and analysis techniques for hybrid multi-rate control systems have then received a great deal of attention from the control community in recent decades \cite{Apostolakis_ACC_1990,ITOP_IMA_journal_1994,Longhi94,VanHaren_CDC_2022}.

Many contributions in this field are based on time or frequency lifting techniques which, given a few assumptions on the sampling rates, allow the multi-rate system to be rewritten as an augmented single-rate model on which classical analysis techniques (such as gain or phase margins evaluations) become applicable \cite{Apostolakis_ACC_1990,ITOP_IMA_journal_1994,VanHaren_CDC_2022}. More precisely, it can be observed that a multi-rate sampled-data system can be first rewritten as  a Linear Periodically Time Varying System (LPTVS) \cite{Bitanti_book_2009} and next transformed into an equivalent single-rate LTI model with expanded inputs and outputs. Such methods are not directly compatible, however, with the introduction of parametric uncertainties in a suitable way for $\mu$ or $IQC$ based robustness analysis techniques, which rely on LFT modeling. An alternative approach is then developed in this paper to generate a single-rate full discrete time LFT model from the initial uncertain hybrid system. Given an uncertain continuous-time linear process in feedback with a discrete-time multi-rate controller, the proposed approach consists of the following steps:
\begin{enumerate}
\item zero-order-hold (ZOH) LFT-based discretization of the continuous-time model at the sampling rate of the fastest control loop,
\item single-rate closed-loop LFT generation by iterative closed-loop evaluations and LFT-preserving down-sampling operations from the fastest to the slowest control loop.
\item application of advanced $\mu$-analysis tools\footnote{This evaluation can be performed either directly in the discrete-time domain for which $\mu$-based tools exist or can be adapted. Alternatively, a bilinear transformation (preserving both stability and the ${\cal H}_\infty$ norm) can be applied to the discrete-time LFT so that more standard continuous-time $\mu$ tools become directly applicable.} for robust stability and ${\cal H}_\infty$ performance evaluation. 
\end{enumerate}

One of the key issues in the above procedure is the LFT-based discretization phase. As is clarified in the following sections, the exact ZOH discretization which must be considered in our context introduces non-rational terms (matrix exponentials) in the model. The LFT structure is then lost, and an important contribution of this paper is to show how a discrete-time LFT model can be obtained at the price of a quantified approximation error.  It is also clarified that, under certain conditions, down-sampling operations preserve the LFT structure at the price of increased complexity.

The remainder of the paper is organized as follows. An overview of the modeling process is given in section II, which also clarifies the class of hybrid multi-rate systems under consideration. Next, the central results of the paper regarding LFT preserving discretization and down-sampling techniques are discussed in section III. Then, $\mu$-based robust performance is briefly detailed for the considered problem in section IV and an easily reproducible illustrative example is presented in section V. Final comments and future directions conclude the paper.

\noindent {\it Notation.} Given two operators $M$ and $N$ where $M$ is block-partitioned in a compatible way with the dimensions of $N$,   the lower-LFT (Linear Fractional Transformation), when it exists, is defined as ${\mathcal F}_l(M,N)=M_{11}+ M_{12} N(I-NM_{22})^{-1} M_{21}$. Similarly, for a possibly different suitable partition, the upper-LFT is defined as ${\mathcal F}_u(M,N)=M_{22}+ M_{21} N(I-NM_{11})^{-1} M_{12}$. Given any structure $\mathbf{X}$ whose elements are real/complex valued matrices or vectors, ${\cal B}_\mathbf{X}$ denotes either the unit ball $\{ X \in \mathbf{X} / \overline{\sigma}(X) \leq 1\}$ or the hypercube $\{ X \in \mathbf{X} / |X_i|  \leq 1\}$. 

\section{Overview of the Modeling Process}

\begin{figure}[thpb]
\centering
\includegraphics[width=0.5\textwidth]{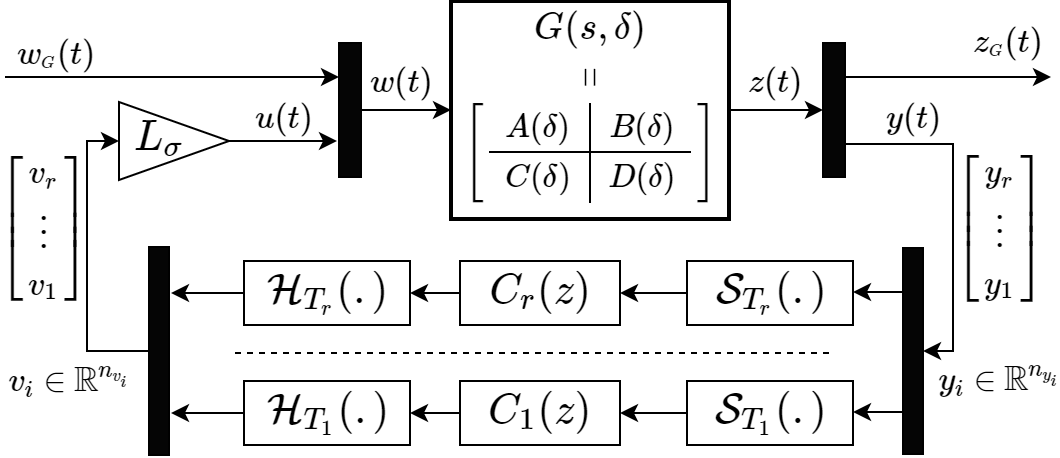}
\caption{Uncertain hybrid continuous/discrete multi-rate closed-loop system}
\label{fig1}
\end{figure}

Consider the class of uncertain hybrid continuous/discrete time multi-rate systems illustrated by the closed-loop diagram in Fig.~\ref{fig1}, where the state-space matrices of the $n^{th}$ order linear continuous-time uncertain process 
\beql{eqssu}
G(s,\delta) : \left\{
\begin{array}{l}
\dot{x} = A(\delta) x + B(\delta) w \\
z = C(\delta) x + D(\delta) w
\end{array}
\right.
\eeq
are assumed to depend \textbf{rationally}\footnote{This assumption is required for LFT modeling, but is not so restrictive in practice since any continuous nonlinear function assumes rational  approximations on a bounded set.} on the vector of normalized parametric uncertainties $\delta = [\delta_1,\ldots,\delta_{n_\delta}]^T \in {\cal B}_\delta$.

Closed-loop stability and performance -- evaluated through the ${\cal H}_\infty$ norm of the transfer from the exogenous inputs $w_G$ to the exogenous outputs $z_G$ -- are ensured by a multi-rate controller involving $r$ loops. As shown in Fig.~\ref{fig1}, each (possibly multivariable) loop operates at a specific frequency $1/T_i$ verifying Assumption \ref{A1}, and is composed of the  following three elements in cascade:
\begin{itemize}
\item ${\cal S}_{T_i}(.)$ : $n_{y_i} \times n_{y_i}$ sampler  where $n_{y_i}$ represents the size of the vector $y_i$ of measurements updated every $T_i$ seconds. Note that $\{y_i\}_{i=1,\ldots, r}$ form a partition of $y$. Then we have $\sum_{i=1}^r  n_{y_i} = p$.
\item $C_i(z)$ :  multivariable $n_{v_i} \times n_{y_i}$ discrete-time controller  with the same period $T_i$ as above.
\item ${\cal H}_{T_i}(.)$ : $n_{v_i} \times n_{v_i}$ zero-order-hold operator.
\end{itemize}

\begin{Assumption}
\label{A1}
The sampling periods $\{T_i\}_{i=1,\ldots, r}$ are assumed to verify $T_{i+1} = q_i T_i$ with $q_i \in \mathbb{N}$.
\end{Assumption}

Each loop finally delivers a continuous-time signal $v_i(t)$ from which the  global control input $u(t)$ is obtained as the output of a linear static gain $L_\sigma \in \mathbb{R}^{m \times \sum n_{v_i}}$. Note that this matrix operator is designed to perform elementary operations (addition, subtraction) on the signals $v_i$ of compatible dimensions. It will then be essentially composed of $0$, $+1$, or $-1$.

\begin{figure}[thpb]
\centering
\includegraphics[width=0.45\textwidth]{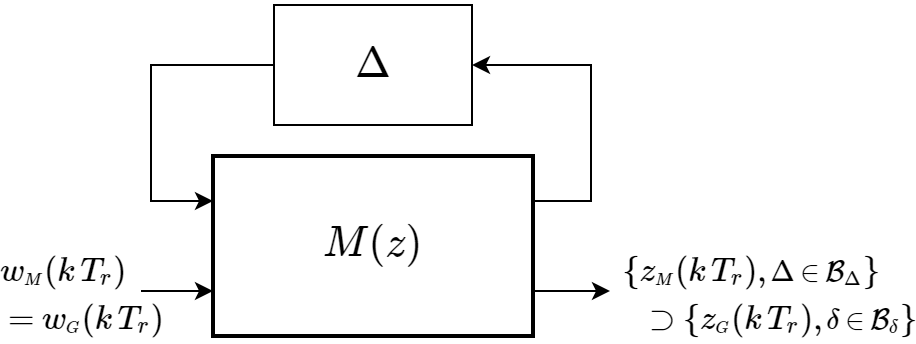}
\caption{LFT-based single-rate full discrete-time approximation}
\label{fig2}
\end{figure}

The main objective of the modeling process is to convert the uncertain hybrid closed-loop model of Fig.~\ref{fig1} into a single-rate discrete-time LFT as illustrated in Fig.~\ref{fig2}, where
\begin{itemize}
\item $M(z)$ denotes an LTI discrete-time interconnection operating at the slowest rate $1/T_r$,
\item $\Delta$ captures all the parametric uncertainties $\delta$ from the continuous-time process and also incorporates discretization errors such that, for any step input profile $w_G(t)$ with constant values $w_G(kT_r)$ on each interval $[kT_r , (k+1) T_r[$, the sampled trajectories $z_G(kT_r)$ of the hybrid multi-rate system are covered by those of  the single-rate and full discrete-time version $z_M(kT_r)$.
\end{itemize}\vspace*{-2mm}

To achieve this goal, the procedure described in Algorithm \ref{algo1} is proposed. It involves two key technical steps (LFT preserving discretization and down-sampling), summarized by relations (\ref{LFTd}) and (\ref{DS}) further discussed in section \ref{secLFTDDS}.
\begin{algorithm}
\caption{Single-rate discretization of an LFT-based uncertain hybrid multi-rate system}
\label{algo1}
\begin{algorithmic} 
\REQUIRE Uncertain hybrid closed-loop system of Fig. \ref{fig1}
\ENSURE Single-rate discrete-time model of Fig. \ref{fig2}
\STATE \textbf{Initial step} Perform a ZOH discretization (see \ref{ZOHD}) of the uncertain continuous-time  model $G(s,\delta)$ with sampling period $T_1$ and generate a discrete-time LFT:
\beql{LFTd}
G(s,\delta) \rightarrow {\cal F}_u(H(z),\Delta_H)
\eeq
Set $\Delta_1=\Delta_H$ and close the fastest loop with sampling period $T_1$ to generate $M_1(z)$ by evaluating the lower-LFT:
\beq 
M_1(z) = {\cal F}_l(H(z),C_1(z))
\eeq
  
\FOR{$i=1, \ldots, r-1$}
\IF{$q_i > 1$}
\STATE  Down-sample the upper-LFT from $T_i$ to $T_{i+1} \hspace*{-1mm}= \hspace*{-1mm}q_iT_i \, $:
\beql{DS}
{\cal F}_u(M_i(z),\Delta_{i})  \rightarrow {\cal F}_u(M_{i+1}(z),\Delta_{i+1})
\eeq
\ELSE
\STATE Set $M_{i+1}(z) = M_i(z)$ and $\Delta_{i+1}=\Delta_i$
\ENDIF
\STATE Update $M_{i+1}(z)$ by closing the loop with $C_{i+1}(z)$:
\beq
M_{i+1}(z) \leftarrow {\cal F}_l(M_{i+1}(z),C_{i+1}(z))
\eeq
\ENDFOR
\STATE \textbf{Final step:} Set $M(z)=M_r(z)$ and $\Delta = \Delta_r$

\end{algorithmic}
\end{algorithm}

\section{LFT preserving ZOH Discretization \&  Down-Sampling techniques}
\label{secLFTDDS}

\subsection{ZOH discretization of a continuous-time LFT model}
\label{ZOHD}
As detailed in \cite{Toth_TCST_2012}, there exist many different techniques to discretize a continuous-time LFT model. Based on a bilinear transformation of the Laplace variable $s$, the well-known Tustin's method is very interesting as it fully preserves the LFT structure without any augmentation of the $\Delta$ block. Unfortunately, this approach is not suitable here, as the discrete-time output signals are not guaranteed to match the continuous-time signals at the sampling times ($z_k \neq z(kT)$). A possible alternative would be a "full ZOH" discretization of $M_G(s)$ where $G(s,\delta)={\cal F}_u(M_G(s),\Delta_G(\delta))$. However, as observed in \cite{Toth_TCST_2012}, this approach becomes inexact as soon as $\delta \neq 0$ and the only rigorous ZOH discretization of (\ref{eqssu}) reads
\beql{eqSSD0}
\left[
\begin{array}{c}
x_{k+1} \\
z_k
\end{array}
\right]
=  {\cal H}_0(\delta)
\left[
\begin{array}{c}
x_{k} \\
w_k
\end{array}
\right]
\eeq
with
\beql{eqM0}
{\cal H}_0(\delta) =
\left[
\begin{array}{cc}
e^{A(\delta)T}  &  \int_0^T e^{A(\delta)\tau} B(\delta) d\tau \\
 C(\delta) & D(\delta)
 \end{array} \right]
\eeq

Interestingly, for any step input signal $w(t)=w_k$, with $t \in [kT, (k+1)T[$, the above expression ensures that the outputs of the continuous-time system (\ref{eqssu}) match exactly those of (\ref{eqSSD0}) at every sampling time: $\forall k \geq 0, z(kT)=z_k$.
However, the presence of matrix exponentials unfortunately destroys the rational dependence on the uncertain parameters $\delta$. Consequently, an LFT model cannot be derived directly from (\ref{eqSSD0}), which must  first  be rewritten using polynomial or rational approximations of the exponentials.

\vspace*{2mm}

\subsubsection{\textbf{First-order rational approximation}}

A commonly used rational approximation of the matrix exponential is based on the first-order Padé approximant:
\beql{pade1}
e^{A(\delta)T} \approx \left(I-\frac{T}{2}A(\delta)\right)^{-1} \left(I+\frac{T}{2}A(\delta)\right) 
\eeq

By introducing a quantification of the error in the above approximation and then substituting the exponentials in (\ref{eqM0}), the following lemma is obtained:

\begin{Lemma}
\label{lem1}
$\forall T \geq 0$
there exists  a bounded positive real $\mu_1(T)$  such that 
\beql{eqH0LFT}
\forall \delta \in {\cal B}_\delta , \;\exists \, \Delta_\epsilon \; / \; {\cal H}_0(\delta) = {\cal F}_u \left(
{\cal H}_1(\delta,\Delta_\epsilon) ,T.I_n \right)
\eeq
\end{Lemma}
with
{\small
\beql{eqH1}
{\cal H}_1(\delta,\Delta_\epsilon) = 
\left[ \hspace*{-1mm}
\begin{array}{ccc}
\frac{1}{2} A(\delta) & (I+\Delta_\epsilon) A(\delta)  & (I-\Delta_\epsilon) B(\delta) \\
I & I & 0 \\
0 & C(\delta) & D(\delta)
 \end{array} \hspace*{-1mm} \right]
\eeq
}

\noindent and
{\small
\beql{eqMb}
\overline{\sigma} (\Delta_\epsilon) \leq \mu_1(T)
\eeq
}

\begin{proof}
Using the power series of the matrix exponential $e^X = \sum_{k=0}^\infty \frac{1}{k!}X^k$, it is readily checked after standard matrix manipulations that (\ref{eqH0LFT}) holds with
\beql{eqDAB}
\Delta_\epsilon = -\frac{T^2 A(\delta)^2}{12}  \underbrace{\sum_{k=0}^\infty \frac{T^k A(\delta)^k}{(1+\frac{k}{2})(1+\frac{k}{3})k!}}_{ \xrightarrow[T \to 0]{} I_n}
\eeq
from which the bound $\mu_1(T)$ is obtained as
{\small
\beql{eqrho1}
\mu_1(T) = \frac{T^2}{12 \; }  \max_{\delta \in {\cal B}_\delta}  \overline{\sigma}  \left( \sum_{k=0}^\infty \frac{T^k A(\delta)^{k+2}}{(1+\frac{k}{2})(1+\frac{k}{3})k!} \right)  \xrightarrow[T \to 0]{} 0
\eeq}
If $A(\delta)$ affinely depends on $\delta$ (often verified in practice), a cheap approximation of $\mu_1(T)$ for low values of $T$ reads
\beql{eqrho1c}
\mu_1(T) \approx \frac{T^2}{12 \; }  \max_{\delta \in {V(\cal B}_\delta)}  \overline{\sigma}  \left( A(\delta)^{2}\right) 
\eeq
where $V({\cal B}_\delta)$ denotes the set of vertices of ${\cal B}_\delta$.
\end{proof}

\begin{Corollary}
\label{cor1}
There exists a linear interconnection matrix ${\cal L}_H$ and a block-diagonal uncertain operator 
\beql{Deltad}
\Delta_\delta = \mbox{diag}(\delta_1 I_{n_1},\ldots,\delta_{n_\delta} I_{n_{n_\delta}})
\eeq
such that
\beql{eqH1LFT}
{\cal H}_1(\delta,\Delta_\epsilon) = {\cal F}_u \left({\cal L}_H, \mbox{diag}(\Delta_\delta, \Delta_\epsilon )\right)
\eeq
\end{Corollary}

\vspace*{2mm}

\begin{proof}
As emphasized by equation (\ref{eqH1}), this straightforward consequence of Lemma \ref{lem1} results from the fact that ${\cal H}_1(\delta,\Delta_\epsilon)$ affinely depends on
\begin{itemize}
\item[$(i)$] the approximation error $\Delta_\epsilon$
{\small
\beql{eqH1aff}
{\cal H}_1(\delta,\Delta_\epsilon) = {\cal H}_1 (\delta,{\bf 0}) + 
\left[
\begin{array}{c}
\Delta_\epsilon \\ {\bf 0} \\ {\bf 0}
\end{array}
\right]
  \big[{\bf 0} \;\, A(\delta) \;\, -B(\delta) \big]
\eeq
}
\item[$(ii)$] the state-space matrices  $A(\delta)$, $B(\delta)$, $C(\delta)$, $D(\delta)$ which, themselves, by assumption,  rationally depend on $\delta$. 
\end{itemize}
The construction of ${\cal L}_H$ associated with the block-diagonal uncertainty structure $\Delta_H=\mbox{diag}\left(\Delta_\delta,\Delta_\epsilon \right)$ is easily realized from (\ref{eqH1aff}) with the help of uncertainty modeling tools based on uss\footnote{see: https://www.mathworks.com/help/robust/ref/uss.html } or gss objects \cite{gsst}.
\end{proof}

\begin{figure}[thpb]
\centering
\includegraphics[width=0.45\textwidth]{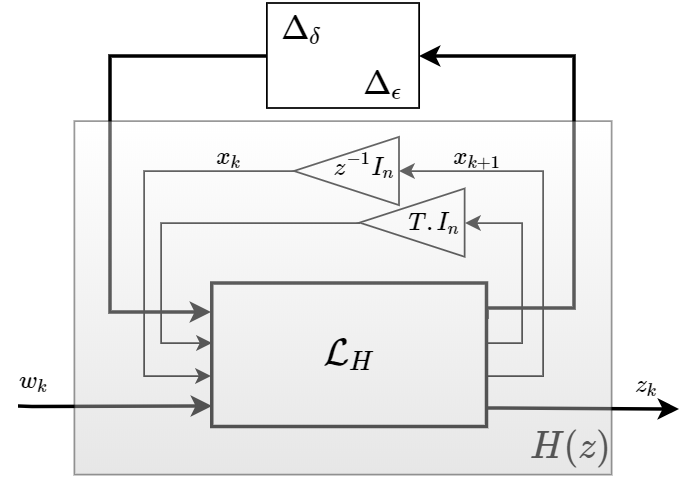}
\caption{A graphical illustration of Lemma \ref{lem1} and its corollary}
\label{fig3}
\end{figure}

\begin{Remark}
Both Lemma \ref{lem1} and its corollary are illustrated by Fig.~\ref{fig3}, where the LFT-based discretization (\ref{LFTd}) introduced in Algorithm \ref{algo1} clearly appears:
\beql{eqHz}
z_k = {\cal F}_u \Big( H(z),\underbrace{\mbox{diag}\left(\Delta_\delta,\Delta_\epsilon \right)}_{\Delta_H}\Big) w_k
\eeq
\end{Remark}

\subsubsection{\textbf{Second-order approximation}}
The upper-bound on the approximation error (\ref{eqMb}) can be drastically reduced by considering higher-order rational approximations of the matrix exponential. Notably, the second-order approximation:
\beql{pade2}
e^{X} \approx \left(I-\frac{X}{2}+\frac{X^2}{12}\right)^{-1} \left(I+\frac{X}{2}+\frac{X^2}{12}\right) 
\eeq

\noindent permits to update equation (\ref{eqH0LFT}) of Lemma \ref{lem1} by replacing ${\cal H}_1(\delta,\Delta_\epsilon)$ with ${\cal H}_2(\delta,\Delta_\epsilon)$ as follows

{\small
\beql{eqH2}
{\cal H}_2(\delta,\Delta_\epsilon) =  
 \left[ \hspace*{-2mm}
\begin{array}{ccc}
(I \hspace*{-2pt} - \hspace*{-2pt} \frac{T}{6}A(\delta)) \frac{A(\delta)}{2} & \hspace*{-6pt} (I \hspace*{-2pt} + \hspace*{-2pt} \Delta_\epsilon) A(\delta)  & \hspace*{-6pt}  (I \hspace*{-2pt} - \hspace*{-2pt} \Delta_\epsilon) B(\delta) \\
I & I & 0 \\
0 & C(\delta) & D(\delta)
 \end{array} \hspace*{-2mm} \right] \;\;\;\;\;\;
\eeq
}

\noindent where $\Delta_\epsilon$ becomes
\beql{eqDE2}
\Delta_\epsilon = -\frac{T^4 A(\delta)^4}{720 \, } \sum_{k=0}^\infty \frac{T^k A(\delta)^k}{(1+\frac{k}{3})(1+\frac{k}{4})(1+\frac{k}{5})k!} 
\eeq
so that the approximated bound (\ref{eqrho1c}) for small values of $T$ and affine parametric dependency becomes
\beql{eqrho2c}
\mu_2(T) \approx \frac{T^4}{720 \; }  \max_{\delta \in {V(\cal B}_\delta)}  \overline{\sigma}  \left( A(\delta)^{4}\right) 
\eeq

In this way, the second-order approximation  saves two orders of magnitude on the norm of $\Delta_\epsilon$. The price to pay is a greater complexity of the LFT model (\ref{eqHz}) with an increase in the size of $\Delta_\delta$ in (\ref{Deltad}), since ${\cal H}_2(\delta,\Delta_\epsilon)$ not only depends on $A(\delta)$ but also on $A(\delta)^2$.

\vspace{1mm}\subsubsection{\textbf{Extension to higher orders}}

Finally, on specific cases involving few uncertainties, higher-order approximations can be considered to reach greater accuracy. Indeed, the upper-bound $\mu_n(T)$ on the $n^{th}$ order approximation error  verifies $\mu_n(T) \propto  \max_{\delta \in {\cal B}_\delta}  \overline{\sigma}  \left( (T.A(\delta))^{2n}\right) $.

\begin{Remark}
Assuming the spectral radius of $T.A(\delta)$ is bounded by $1$ throughout the uncertainty domain (always verified in practice), it is easily checked that  $\overline{\sigma} \left( \Delta_\epsilon \right) \xrightarrow[n \to \infty]{} 0$.
\end{Remark}

\subsection{Down-sampling: an LFT preserving operation}
\label{subDS}
Let us now focus on the down-sampling operation summarized by equation (\ref{DS}) of Algorithm \ref{algo1}, and clarify that the LFT structure of the model is preserved. To do so, the following lemma is introduced.
\begin{Lemma}
\label{lem2}
Consider $F_{\tau}(z)=C(zI-A)^{-1}B+D$, a discrete-time system  with $n$ states, $m$ inputs and sampling period $\tau$. Then, for all $q \in \mathbb{N}$, the down-sampled system with period $q \tau$ is $F_{q\tau}(z)=C(zI-A_q)^{-1}B_q+D$ with
\beql{eqDSF}
[A_q \;\;  B_q] = [I_n  \;\; {\bf 0} ]
\left[
\begin{array}{cc}
A & B \\ {\bf 0} & I_m
\end{array}
\right]^q
\eeq
\end{Lemma}
\vspace*{1mm}
\begin{proof} This standard result is readily obtained by computing $x_{k+q} = A_q x_k + B_q w_k$ from $x_{k+1} = A x_k + B w_k$ and holding $w_k$ constant ($w_k=w_{k+1}=\ldots = w_{k+q-1}$).
\end{proof}
\vspace*{1mm}
Now, following the notation of Algorithm \ref{algo1}, after closing the first $i$ control loops, the system ${\cal F}_u(M_i(z),\Delta_i)$ to be down-sampled from period $T_i$ to $T_{i+1}=q_i T_i$ is an upper-LFT. Then, replacing the fixed matrices $[A \;\,  B]$  in Lemma \ref{lem2} by $[A(\Delta_i) \; B(\Delta_i)]$, equation  (\ref{eqDSF}) can be re-interpreted as the product of $q=q_i$ LFTs. As a result, and further assuming without a significant loss of generality in practice that the $C$ and $D$ matrices are fixed, it is readily checked that $\Delta_{i+1}$ in the down-sampled model ${\cal F}_u(M_{i+1}(z),\Delta_{i+1})$ verifies
\beql{eqDeli}
\Delta_{i+1} = \mbox{diag} \, \big(  \underbrace{\Delta_i, \ldots, \Delta_i}_{q_i \; times} \big )
\eeq
As a result
\beql{eqDel}
\Delta = \mbox{diag} \, \big(  \underbrace{\Delta_H, \ldots, \Delta_H}_{N \; times} \big ) \; \mbox{with}  \; N = \prod_{i=1}^{r-1}q_i
\eeq
\section{On $\mu$-based Robust Performance Analysis}
\label{secmu}
At this point, a standard discrete-time uncertain LFT model $z_M = {\cal F}_u(M(z),\Delta)w_M$ is obtained, whose stability and $\mathcal{H}_\infty$ performance can now be analyzed.
\begin{Remark}
Since both stability and the ${\cal H}_\infty$ norm are preserved under any bilinear transformation $s=k\frac{z-1}{z+1}$, which also preserves the LFT structure without altering the $\Delta$ block, the analysis can be carried out in the continuous-time domain on ${\cal F}_u(P(s),\Delta)$ with $P(s)=M(-\frac{s+k}{s-k})$.
\end{Remark}

Let $\bDelta$ be the set of all matrices with the same structure as $\Delta$ in~(\ref{eqDel}). Recall that all uncertainties are normalized, so the considered uncertainty domain is simply $\BD$. The following two quantities are evaluated using $\mu$-analysis:
\begin{itemize}
\item the robust stability margin:
$$k_{r}=\max\left\{k\ge 0:P(s)-\Delta \textrm{ is stable}\ \forall\Delta \in k\BD \right\}$$
\item the worst-case $\mathcal{H}_\infty$ performance level (if $k_{r}>1$):
$$\displaystyle\gamma_{wc}=\max_{\Delta\in\BD}\|{\cal F}_u(P(s),\Delta)\|_\infty$$
\end{itemize}
The underlying theory is not presented here due to space limitations, but the interested reader can for example refer to~\cite{ZhDoGl96_book,Fe99_book}. Only a few facts are briefly recalled to facilitate the understanding of section~\ref{sec:robanal}. $\mu$-analysis basically consists of computing the peak value over the entire frequency range of the structured singular value $\mu_{\bDelta}$. This computation being NP-hard in general, lower / upper bounds $\underline{\mu}$ / $\overline{\mu}$ are usually determined instead of the exact value, from which bounds $\underline{k}_{r}$ / $\overline{k}_{r}$ on $k_{r}$ and $\underline{\gamma}_{wc}$ / $\overline{\gamma}_{wc}$ on $\gamma_{wc}$ are derived. Much work has been done in the past decades to reduce the gap between these bounds, and (almost) exact values of $k_{r}$ and $\gamma_{wc}$ are now obtained in most cases with a reasonable computational time~\cite{RoBi15_cep}. The main reason why the gap sometimes remains non-negligible and the computational time significant is the presence of uncertainties repeated many times in $\Delta$, which is precisely the case in this paper, see~(\ref{eqDel}). A branch-and-bound algorithm can be used to overcome this issue. The uncertainty domain $\BD$ is cut into smaller and smaller subsets until the relative gap between the highest lower bound and the highest upper bound on $k_{r}$ or $\gamma_{wc}$ computed on all subsets becomes less than a user-defined threshold. This algorithm is known to converge for uncertain systems with only real parametric uncertainties. However, it suffers from an exponential growth of computational complexity as a function of the number of uncertainties. This can be alleviated using the $\mu$-sensitivities, which provide a very efficient way to
detect the most critical uncertainties, and therefore to decide in which directions to cut the uncertainty domain to quickly reduce the gap. This strategy is implemented in the routine \textit{mubb} of the Matlab SMART Library~\cite{Roo13}, which computes tight bounds on $k_{r}$ and $\gamma_{wc}$ with a very reasonable CPU time (see section~\ref{sec:robanal}).

\section{Illustrative Example}
\label{secappli}
\begin{figure}[thpb]
\centering
\includegraphics[width=0.48\textwidth]{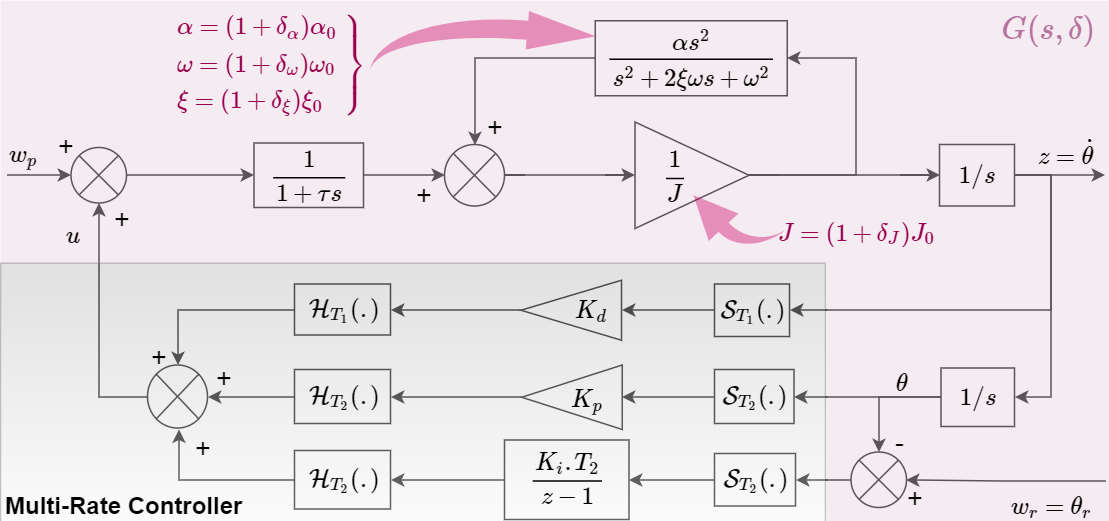}
\caption{Hybrid closed-loop model with a multi-rate PID controller}
\label{fig4}
\end{figure}

The hybrid closed-loop model under consideration is illustrated by Fig.~\ref{fig4}. The continuous plant $G(s,\delta)$ is a double integrator (pure inertia $J=1$) with a positive feedback involving a poorly damped second order transfer function ($\alpha=0.5, \xi=0.001$, $\omega=4$\;rad/s). The actuator is represented by a first-order model with time-constant $\tau=0.2 \,s$. Such a model is commonly used to represent the dynamics of a flexible satellite consisting of a central body and a solar panel. Normalized multiplicative uncertainties $\delta_J$, $\delta_\alpha$, $\delta_\omega$ and $\delta_\xi$ are then applied to the plant parameters to introduce   variations of  $10 \%$ on each of them. The gains $K_p=1.65$, $K_i=0.5$ and $K_d=2.7$ of the PID controller are tuned by a standard continuous-time approach, so that the dominant mode of the nominal closed-loop system has a frequency of 0.5 rad/s. The nominal delay margin associated with this tuning is $\tau_d=0.08\,s$. This PID controller is then digitalized and implemented in a multi-rate setting as shown in Fig.~\ref{fig4}. The derivative loop operates at sampling time $T_1$, while the proportional and integral (for which a forward-Euler approximation is used) loops share the same larger sampling time $T_2 = q T_1$. 

\subsection{Preliminary analysis: a counter-intuitive result?}
\label{subPrelA}
\begin{figure}[thpb]
\centering
\includegraphics[width=0.5\textwidth]{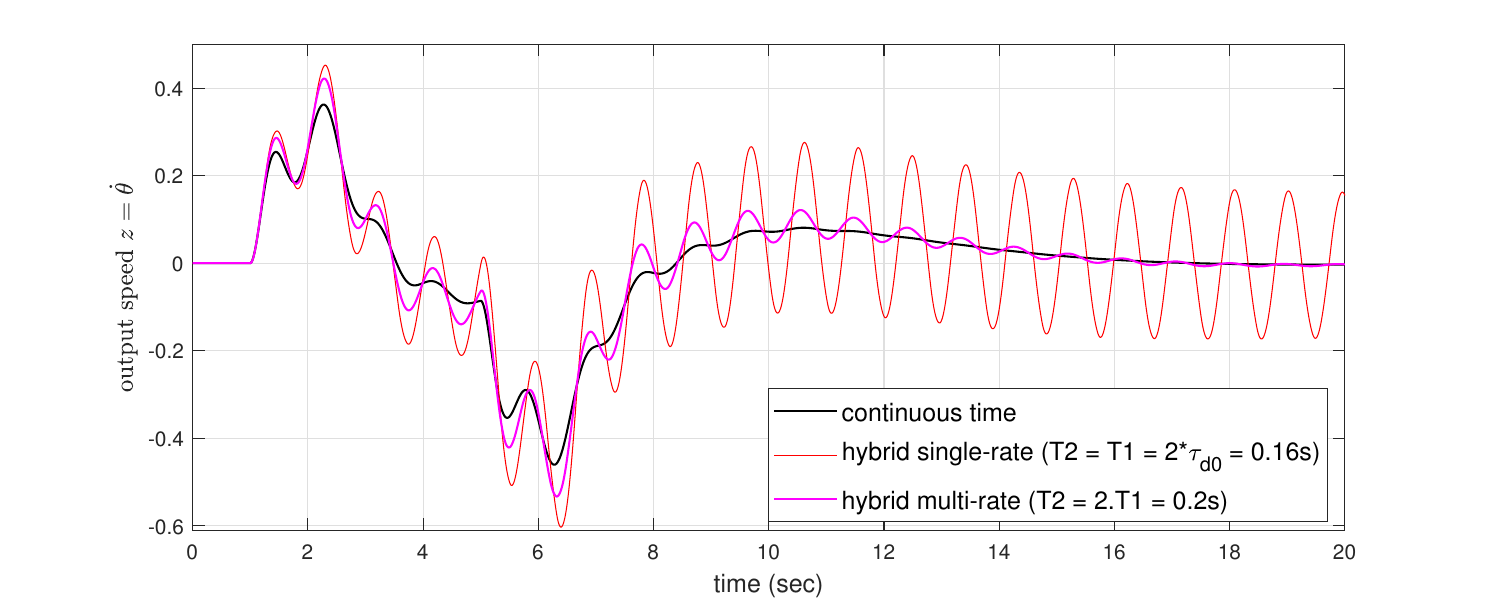}
\caption{\textbf{Nominal configuration}: continuous-time \textit{vs} hybrid single \& multi-rate implementations}
\label{mfig1}
\end{figure}

\begin{figure}[thpb]
\centering
\includegraphics[width=0.5\textwidth]{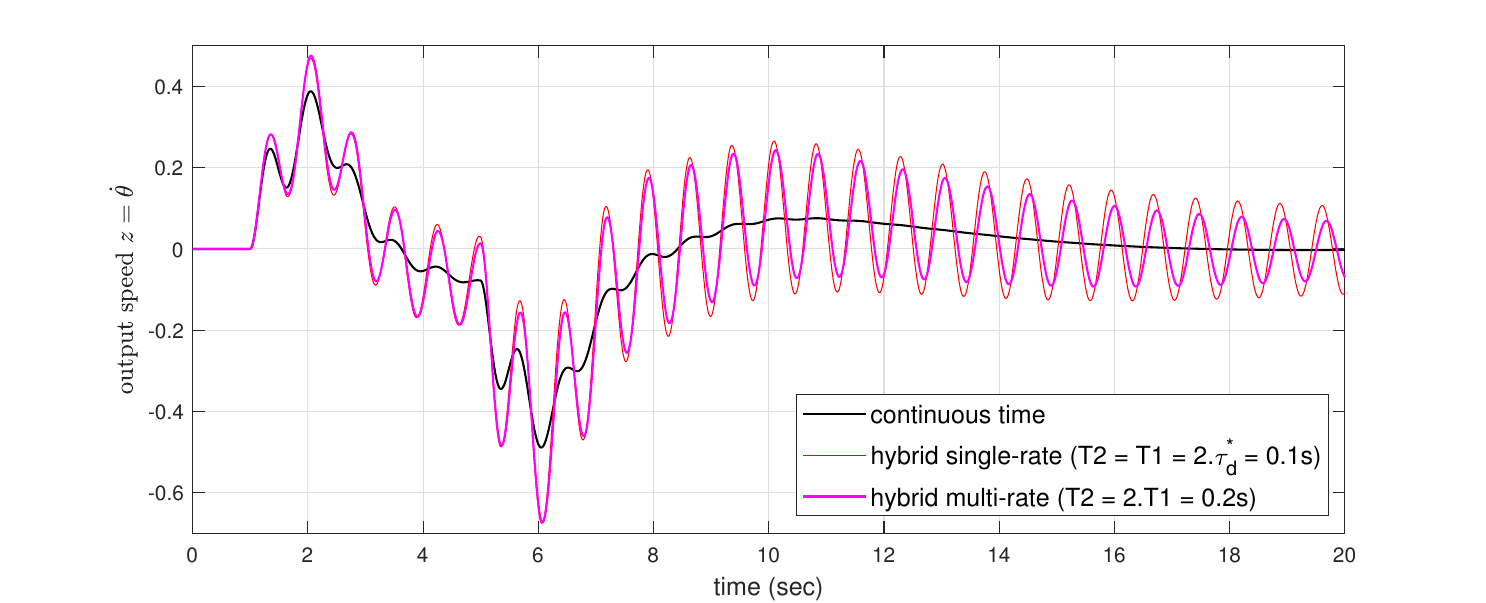}
\caption{\textbf{Critical configuration}: continuous-time \textit{vs} hybrid single \& multi-rate implementations}
\label{mfig2}
\end{figure}

A preliminary analysis is performed by applying a unit step disturbance $w_p$ on the control input signal for 5 seconds. The output signal $z=\dot{\theta}$ is first plotted (Fig.~\ref{mfig1}) in the nominal parametric configuration ($\delta=0$). Next, a critical configuration (maximum values of $\alpha$ and $\omega$, and minimum value of $J$) is displayed in Fig.~\ref{mfig2}. In both cases, the continuous-time reference (black) is compared with two hybrid implementations. The first one (visualized in red) corresponds to a single rate implementation ($T_2=T_1=T$) where $T \approx 2 \tau_d$ is set at almost twice the delay margin (with $\tau_{d_0} \approx 0.08 \,s$ in the nominal case and $\tau_d^* \approx 0.05 \,s$ in the critical case). The second one (magenta plots) implements a multi-rate set-up with fixed sampling times $T_2=2T_1=0.2\,s$ whatever the parametric configuration.


In both configurations, the single-rate simulations (in red) confirm the classical result, which predicts the stability of the hybrid continuous/discrete system as long as the sampling period remains less than twice the delay margin. But, interestingly, as shown in Fig.~\ref{mfig1}, the proportional and integral loops can accommodate a higher sampling period ($T_2 > 2 \tau_d$) by reducing that of the derivative loop ($T_1< 2 \tau_d$). With the proposed tuning $T_2=2T_1=0.2\,s$ the oscillations on $\dot{\theta}$ are strongly attenuated after $15 \, s$.

More surprisingly (\textit{a priori}), in the critical configuration, the multi-rate implementation outperforms the single-rate version, even though in both cases the derivative loop operates at the same rate ($T_1=0.1 \,s$). Thus, by reducing the rate (increasing $T_2$) of the proportional and integral "outer-loops", the stability is improved, which actually makes sense, but is not easily captured by analysis techniques in an uncertain context. This simple example is then a very good benchmark to evaluate the proposed methodology.

\subsection{LFT modeling of the hybrid multi-rate closed-loop model}

The first step is to run Algorithm \ref{algo1}. To do so, the open-loop plant $G(s,\delta) = {\cal F}_u(M_G(s),\Delta_G)$ is first written in the LFT format. A fifth-order ($n_s=5$) model is easily obtained with $\Delta_G = \mbox{diag}(\delta_J,\delta_\alpha,\delta_\omega I_2,\delta_\xi)$. Next, using the rational approximations discussed in subsection \ref{ZOHD}, this model is discretized with $T_1=0.1 \, s$ and the approximation errors are evaluated. Then, the derivative loop is closed, a down-sampling operation from $T_1$ to $T_2=2T_1$ is applied (see subsection \ref{subDS}) and finally the proportional \& integral loops are closed. For comparison purposes, the algorithm is also applied after a Tustin discretization (LFT preserving but unfortunately not suited) or a full ZOH discretization (exact for $\delta=0$, but inexact and even theoretically wrong for $\delta \neq 0$) of $M_G(s)$. The results are summarized in table \ref{tab1}.

\begin{table}[thpb]
\caption{Complexity of ${\cal F}_u(M(z),\Delta)$ \textit{vs} quality of the approximation}
\label{tab1}
\centering
\begin{tabular}{|c|c|c|c|}
\hline
Discretization method & structure of $\Delta_\delta$ & size of $\Delta_\epsilon$ & $\overline{\sigma}(\Delta_\epsilon)$ \\
\hline
Rational approx. (order 1) & $I_2 \otimes \Delta_G$ & $10 \times 10$ & $0.15$ \\
\hline
Rational approx. (order 2)  & $I_4 \otimes \Delta_G$ & $20 \times 20$ & $0.0012$ \\
\hline
Full ZOH  & $I_2 \otimes \Delta_G$ & NA & NA \\
\hline
Tustin  & $I_2 \otimes \Delta_G$ & NA & NA \\
\hline
\end{tabular}
\end{table}

\begin{figure}[thpb]
\centering
\includegraphics[width=0.5\textwidth]{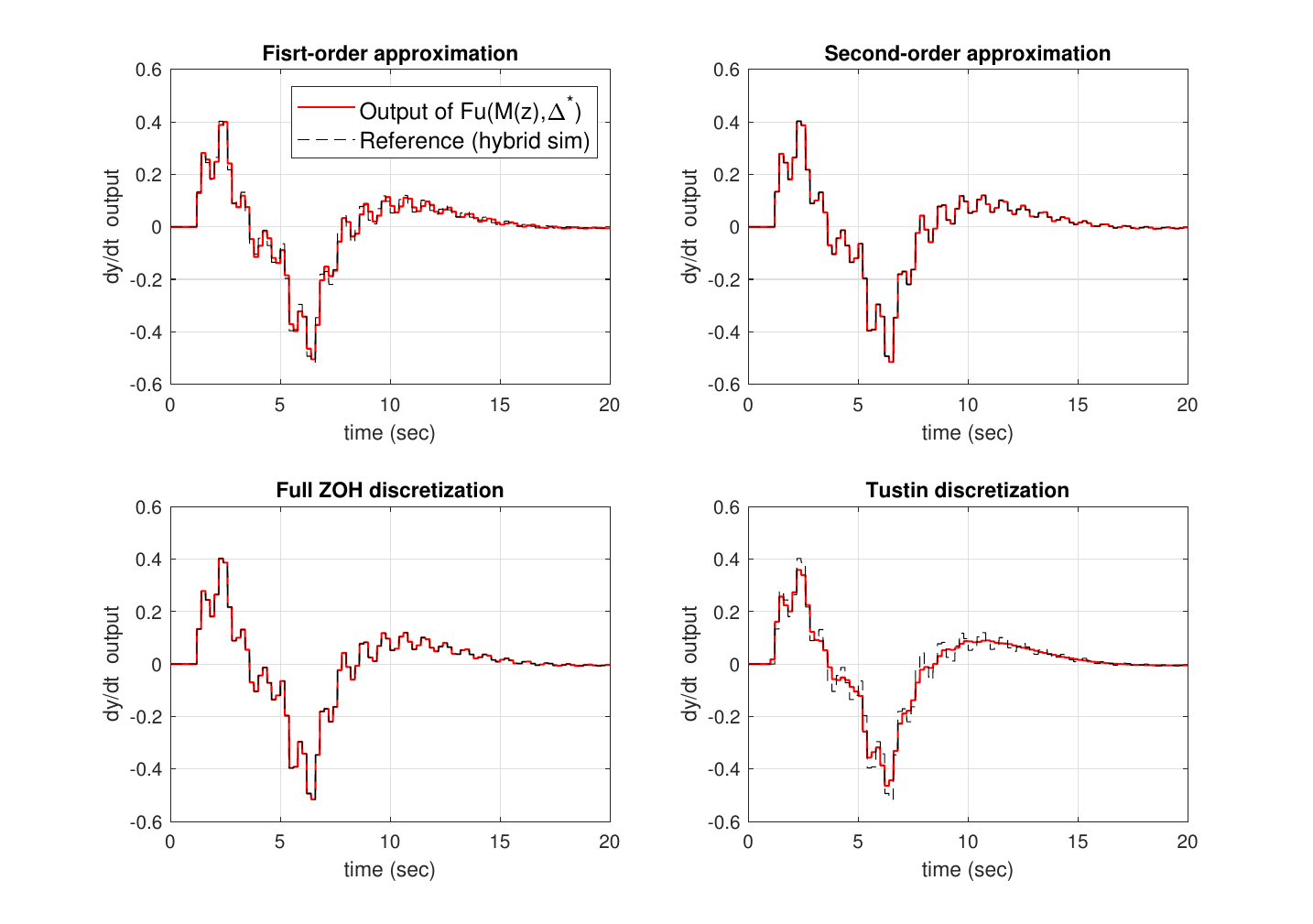}
\caption{\textbf{Nominal configuration}: hybrid multi-rate vs ${\cal F}_u(M(z),0)$}
\label{mfig3}
\end{figure}

\begin{figure}[thpb]
\centering
\includegraphics[width=0.5\textwidth]{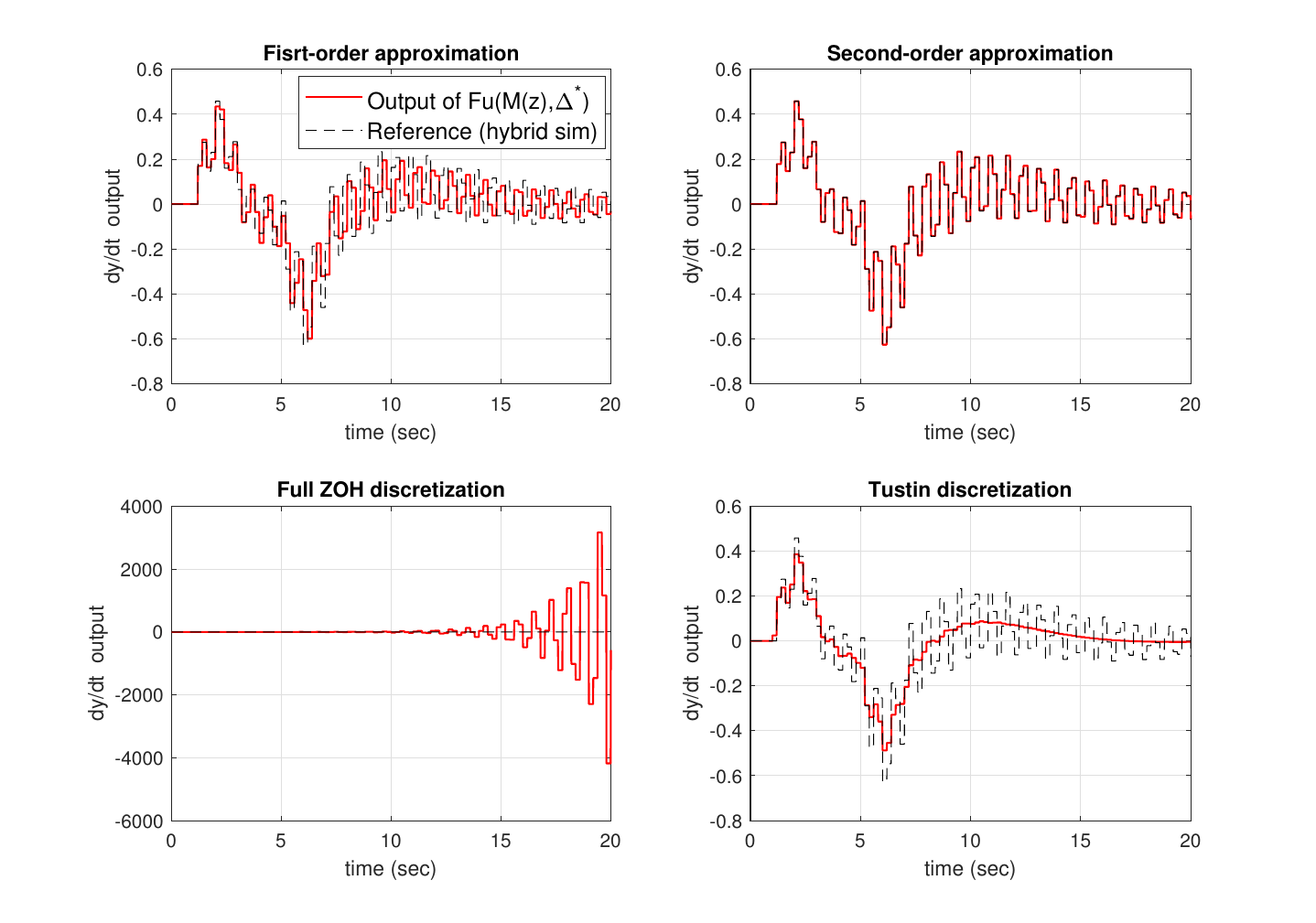}
\caption{\textbf{Critical configuration}: hybrid multi-rate vs ${\cal F}_u(M(z),\Delta^*)$}
\label{mfig4}
\end{figure}

For each of the above four models, the simulations of subsection \ref{subPrelA} are now replayed in the nominal case ($\delta=0$) and for the same critical parametric configuration. As can be seen from the lower left subplot in Fig.~\ref{mfig3}, the method based on the full ZOH discretization exactly reproduces the output of the hybrid multi-rate system. This result was expected, since this discretization is exact without uncertainty. However, this is  no longer the case in the critical configuration, where severe instability appears as revealed by the same subplot in Fig.~\ref{mfig4}. 

Conversely, the Tustin-based approach (displayed on the lower-right subplots) introduces strong distortions whatever the configuration and fails to capture the oscillatory behavior (notably in the critical case). 

Finally, for both configurations, the two closed-loop LFT models (see upper subplots in Fig.~\ref{mfig3} and \ref{mfig4}) generated from the rational approximations of subsection III.A produce good (even excellent with the second-order approximation) results. 

\begin{Remark}
\label{remappr2}
Despite a higher complexity, the second-order approximation offers a major advantage in this application. As shown by the two upper-right subplots, and confirmed by the very small value of  $\overline{\sigma}(\Delta_\epsilon) \leq 1.2 \times 10^{-3}$, the accuracy of the associated LFT is indeed extremely high. 
\end{Remark}

\subsection{Robust stability \& performance analysis}\label{sec:robanal}

As discussed in section \ref{secmu}, a bilinear transformation is finally applied to the above discrete-time LFT models, where $\Delta_\epsilon$ is also introduced to obtain guaranteed results. In the general case, from equations (\ref{eqDAB}) or (\ref{eqDE2}), the simplest way would be to consider $\Delta_\epsilon$ as a real non-structured full block, for which specific scaling operators should be introduced in the $\mu$ upper-bound characterization. In this application however, due to the particular structure of the matrix $A(\delta)$, there are many zeros in $\Delta_\epsilon$. This property can be exploited to represent the approximation error by a limited number of scalar parametric uncertainties, which considerably reduces the conservatism. 

\subsubsection{\textbf{Robust stability}}

Upper and lower bounds on the robust stability margin $k_r$ are first evaluated via a standard call to the \textit{mubb} routine (see section \ref{secmu}). A first-order rational approximation (\#1) without and with modeling errors is initially considered. Since the bound on $\Delta_\epsilon$ remains rather high in this case, robust stability cannot be ensured on the entire uncertainty domain ($\underline{\mu}=1/\overline{k}_r=2.31 >1$). In the second-order case (\#2), however, robust stability is easily ensured with $\overline{\mu}=1/\underline{k}_r=0.86 <1$. Moreover, this result is obtained with a very limited number of iterations of the branch-and-bound algorithm, which explains why only $5 \,s$ are required despite a more complex model. Note also that the conservatism introduced with the modeling error is very low ($\overline{\mu}=0.86$ compared to 0.84 without the error), this time since the bound on $\Delta_\epsilon$ is very small.

\begin{table}[thpb]
\caption{Multi-rate robust stability analysis results}
\label{tab2}
\centering
\begin{tabular}{|c|c|c|c|}
\hline
Model type & $\underline{\mu}_{\ }$ & $\overline{\mu}$ & CPU time\footnote{All computations were performed on a standard laptop with i5-3GHz processor and 16 Go RAM installed} \\
\hline
Rational \#1 $(\Delta_\epsilon=0)$ &  0.79 & 0.83 & 5 s \\
\hline
Rational \#1 ($(\Delta_\epsilon \neq 0)$ &  2.31 & 2.43 & 27 s \\
\hline
Rational \#2 $(\Delta_\epsilon=0)$ &  0.80 & 0.84 & 3 s \\
\hline
\textbf{Rational \#2 $(\Delta_\epsilon \neq 0)$} &  \textbf{0.82} & \textbf{0.86} &\textbf{ 5 s} \\
\hline
Full ZOH &  2.41 & 2.53 & 0.2 s \\
\hline
Tustin &  0.30 & 0.31 & 113 s \\
\hline
\end{tabular}
\end{table}

For comparison purposes, robust stability analysis is also applied to the models generated from the full ZOH and Tustin discretizations. In the first case, with $\overline{\mu}=2.53$, stability is only proved for $4\%$ variations of the parameters, while in the second case, with $\overline{\mu}=0.31$, stability remains guaranteed much beyond the parametric domain, up to $33\%$ variations. These results confirm the time-domain simulations in Fig.~\ref{mfig4} and the invalidity of these two models.

\subsubsection{\textbf{Robust performance}}

Robust performance is now considered through the evolution of the ${\cal H}_\infty$ norm of the transfer ${\cal T}_{w_p \rightarrow z}(s)$ on the uncertainty domain. Upper and lower bounds on the worst-case ${\cal H}_\infty$ performance level $\gamma_{wc}$ are computed for four different models generated from the second-order approximation without ($\Delta_\epsilon=0$) and with ($\Delta_\epsilon\neq 0$) modeling error. The first two models (denoted MR) correspond to the multi-rate case ($T_2=2T_1=0.2 \, s$) while the last two (denoted SR-HF) correspond to a single-rate configuration ($T_2=T_1=0.1 \, s$) \textbf{at the highest frequency}. 

\begin{table}[thpb]
\caption{Robust ${\cal H}_\infty$ performance analysis: multi-rate \textit{vs} single-rate}
\label{tab3}
\centering
\begin{tabular}{|c|c|c|c|}
\hline
Model type & $\underline{\gamma}_{wc_{\ }}$ & $\overline{\gamma}_{wc}$ & CPU time \\
\hline
MR / Rational \#2 $(\Delta_\epsilon=0)$ &  6.71& 6.91 & 40 s \\
\hline
\textbf{MR / Rational \#2 $(\Delta_\epsilon \neq 0)$} &  \textbf{7.12} & \textbf{8.55} &\textbf{ 100 s} \\
\hline
SR-HF / Rational \#2 $(\Delta_\epsilon=0)$ &  10.48& 11.00 & 50 s \\
\hline
SR-HF / Rational \#2 $(\Delta_\epsilon \neq 0)$ &  11.44 & 12.01 & 37 s \\
\hline
\end{tabular}
\end{table}

Very interestingly, a significant increase (beyond $40 \%$) of the two bounds on the worst-case ${\cal H}_\infty$ norm can be observed when the control loops all operate at the highest frequency. This result confirms the preliminary analysis and the fact that the flexible mode is more rapidly damped (whatever the parametric configuration) when the integral and proportional loops operate at a slower rate.



\section{Conclusions and Future Works}

A new approach has been proposed in this paper to describe an uncertain hybrid and multi-rate closed-loop system by a discrete-time and single-rate LFT model. Moreover, modeling errors have been quantified, allowing them to be integrated into a guaranteed $\mu$-based robustness analysis framework. Both the modeling process and the robustness analysis have been successfully evaluated on an easily reproducible and realistic benchmark.

Future work will be devoted to more specific adaptations of the robustness analysis tools to this new context of discrete-time and multi-rate systems. Notably, the case of the robust ${\cal H}_2$ norm, whose value is not preserved by the bilinear transformation, should be carefully investigated. This metric is indeed of high practical interest to quantify pointing errors in space-oriented control applications.

\end{document}